\documentclass[aps,prl,twocolumn,amsmath,amssymb,showpacs]{revtex4}
\usepackage{graphicx}
\usepackage{amsmath}
\usepackage{bm}
\begin{document}

\title{Quantum limit for measurement of a weak classical force coupled to a noisy quantum-mechanical oscillator} 

\author{C. L. Latune}
\affiliation{Instituto de F\'\i sica, Universidade Federal do Rio de Janeiro, 21941-972 Rio de Janeiro, Brazil}
\author{B. M. Escher}
\affiliation{Instituto de F\'\i sica, Universidade Federal do Rio de Janeiro, 21941-972 Rio de Janeiro, Brazil}
\author{R. L. de Matos Filho}
\affiliation{Instituto de F\'\i sica, Universidade Federal do Rio de Janeiro, 21941-972 Rio de Janeiro, Brazil}
\author{L. Davidovich}\email{ldavid@if.ufrj.br}
\affiliation{Instituto de F\'\i sica, Universidade Federal do Rio de Janeiro, 21941-972 Rio de Janeiro, Brazil}
\date{\today}
\begin{abstract}
Precise measurements of tiny forces and displacements play an important role in science and technology. The precision of recent experiments, while beginning to reach the limits imposed by quantum mechanics, is necessarily spoiled by the unavoidable influence of noise. Here we obtain a quantum limit for the uncertainty in the estimation of a resonant classical force acting on a noisy quantum-mechanical oscillator. We determine the best initial state of the oscillator and the best measurement procedure, thus getting a rigorous and useful benchmark for experiments aiming to detect extremely small forces. 
\end{abstract}
\maketitle

Several areas of science and technology rely on the capacity of measuring  tiny forces and displacements \cite{brag0,cavesrmp,brag,binnig,bocko,mamin,knobel,poggio}.  The sensitivity of recent experiments has attained extremely high levels \cite{regal,maiwald,teufel,biercuk}, so that the limits imposed by quantum mechanics start playing an important role. The  measurements are affected however by the unavoidable interaction between these systems and the environment. Evaluating this effect is  a difficult task, since the determination of the ultimate precision limit in the presence of noise is still a challenging problem in quantum mechanics. Here we determine the quantum limit for the uncertainty in the estimation of a weak resonant classical force acting on a noisy quantum-mechanical oscillator, which is used as a probe for the measurement of the force, through a sequence of discrete measurements.  Our solution leads to the best probe state, for a given average energy of the oscillator, and to a measurement procedure that attains, in the asymptotic high-energy limit, the so-called ``potential sensitivity,'' which defines a level of sensitivity in the estimation of a force in the presence of thermal noise that cannot be surpassed by any measurement strategy~\cite{brag}. As a specific application, we derive a precise lower bound for the uncertainty in the estimation of the force acting on a trapped ion under realistic experimental noise conditions~\cite{maiwald}.

The detection of a weak classical force is an example of the general problem of parameter estimation. A typical procedure  consists in sending a probe in a known initial state through some parameter-dependent physical process, and measuring the final state of the probe, estimating then from this measurement the value of the parameter. The precision generally depends on the initial state of the probe, on the measurement and estimation procedure, on the dynamical process, and on the amount of resources used in the measurement (quantified for instance by the number of probes or the energy of each probe). It is given by the Cr\'amer-Rao bound \cite{Cramer1946,rao}, which relates the uncertainty $\delta x$ in the estimation of a parameter $x$ to the Fisher information \cite{fisher2}, ${\cal F}(x)$, defined in terms of the conditional probability density $p(\xi|x)$ of getting the outcome $\xi$ of the measurement when the value of the parameter is $x$:
\begin{equation}\label{cr}
\delta x=\sqrt{\langle\left(x_{\rm est}-x_{\rm true}\right)^2\rangle}\ge {1\over \sqrt{\nu {\cal F}(x_{\rm true})}}\,,
\end{equation}
where 
\begin{equation}\label{fisher}
{\cal F}(x)=\int d\xi\left\{ {1\over p(\xi|x)}\left[{\partial p(\xi|x)\over \partial x}\right]^2\right\}\,,
\end{equation}
$x_{\rm est}$ is the estimated value of the parameter for a possible measurement result, $x_{\rm true}$ is the true value of the parameter, $\nu$ is the number of repetitions of the experiment and the average in (\ref{cr}) is taken over all possible measurement results. The above expression holds for unbiased estimators, for which $\langle x_{\rm est}\rangle=x$. In general, the lower bound in (\ref{cr}) is tight for $\nu\rightarrow\infty$, however  if $p(\xi|x)$ is Gaussian the bound is attainable for any $\nu\ge1$  \cite{Cramer1946,rao,fisher2}. Better precision is obtained upon increasing the Fisher information or the number of repetitions.

Quantum mechanics imposes restrictions on the precision of the estimation, since two outgoing states corresponding to two different values of the parameter are not necessarily distinguishable, and measurements must conform to quantum constraints. On the other hand, quantum features, like entanglement and squeezing, help to increase the estimation accuracy beyond the standard limits~\cite{bollinger,Braunstein1996,Huelga1997,Dowling1998,Lee2002,Giovannetti2004,Giovannetti2006,Huver2008}, yielding better precision for the same amount of resources. 

The application of the Cr\'amer-Rao bound to quantum theory was initiated by Helstrom \cite{Helstrom1976} and Holevo~\cite{Holevo1982}.  Braunstein and Caves \cite{Braunstein1994} showed that maximization of the Fisher information in (\ref{fisher}) over all possible measurement procedures, leading to the so-called quantum Fisher information, yields, in the noiseless case, a simple expression for the corresponding uncertainty (\ref{cr}), which has been applied to many different systems \cite{Braunstein1996,Giovannetti2006,GLM2011:222}. 
If the initial state of the probe is a pure state $|\psi_0\rangle$ and the state of the outgoing probe is $|\psi(x)\rangle=\hat U(x)|\psi_0\rangle$, where $\hat U(x)$ is an $x$-dependent unitary operator, then the corresponding quantum Fisher information is four times the variance, calculated in the state $|\psi_0\rangle$,  of the operator
 \begin{equation}\label{h}
\hat{\cal H}(x):=i{d\hat U^\dagger(x)\over dx}\hat U(x)\,,
\end{equation}
that is,
\begin{equation}\label{fq}
{\cal F}_Q(x)=4\langle\psi_0|[\Delta\hat {\cal H}(x)]^2|\psi_0\rangle\,.
\end{equation}

However, the estimation of parameters in the presence of noise poses formidable challenges. Only for very special situations it is possible to derive analytic lower bounds for the uncertainty \cite{Sarovar2006,monras, Fujiwara2008,walmsleyprl,rafal}. 

A practical procedure for evaluating the precision of estimation was introduced in \cite{escher,escherbjp} and further developed in \cite{nicim}.  Here we show that this method leads to a tight analytical bound for the uncertainty in the estimation of a resonant classical force acting on a noisy quantum oscillator with a fixed average energy, considered as a resource. 

We start by considering momentum measurements on the oscillator, for an initial Gaussian state. We show next that the corresponding Fisher information actually coincides with the quantum Fisher information for estimation of the force, with the same initial states. Furthermore, under the constraint of fixed initial average energy of the harmonic oscillator, we show that squeezed Gaussian states maximize the quantum Fisher information, thus yielding the ultimate precision limit for the estimation of the force.

{\it Momentum measurements on the noisy harmonic oscillator}. The Hamiltonian of a harmonic oscillator under the action of a resonant classical force is given, in terms of dimensionless variables, by $\hat H_S/\hbar\omega=(\hat P^2+\hat X^2)2+F\cos(\omega t)\hat X$, where, in terms of the momentum $\hat p$, the position $\hat x$, the mass $m$, the oscillation frequency $\omega$, and the force amplitude $f$, the dimensionless variables are defined by $\hat P=\hat p/\sqrt{m\hbar\omega}$, $\hat X=\hat x\sqrt{m\omega/\hbar}$, so that $[\hat X,\hat P]=i$, and $F= f/\sqrt{\hbar m\omega^3}$. The aim here is the estimation of $f$. We note that the resonance condition is the most favorable situation for estimating the force.

Setting $\hat{X} = (\hat{a}^{\dag} + \hat{a})/\sqrt{2}$ and $\hat{P} = i(\hat{a}^{\dag} - \hat{a})/\sqrt{2}$, so that $[\hat a,\hat a^\dagger]=1$, we have, in the interaction picture,
$\hat H_I(t) =  \hbar\omega F\cos(\omega t) (\hat{a} e^{-i \omega t} + \hat{a}^{\dagger} e^{i \omega t})/\sqrt{2}$, with $\hat H_0 = \hbar \omega/2 (\hat{P}^2 + \hat{X}^2)=\hbar\omega\left(\hat a^\dagger\hat a+{1\over2}\right)$. In the rotating-wave approximation, valid for $\omega t\gg1$, one neglects oscillating terms, getting then $\hat H_I(t)=\hbar\omega F\hat X/2$, corresponding to a nunitary evolution implemented by the momentum displacement operator $\hat U(t)=\exp(-i\omega t F\hat X/2)$.

Physical insight into the dynamics of the forced harmonic oscillator in the presence of thermal noise can be obtained from the corresponding Heisenberg-Langevin equation for the momentum operator:   
\begin{equation}\label{ptgamma}
{d\hat P/ dt}=\omega F/2-\gamma\hat P/2+  \hat f_\gamma(t)\,,
\end{equation}
where $\gamma$ is the friction coefficient and $\hat f_\gamma(t)$ is a Hermitian fluctuation force, with $\langle \hat f_\gamma(t)\rangle=0$ and $\langle\hat f_\gamma(t) \hat f_\gamma(t^\prime)\rangle=\gamma(n_T+1/2)\delta(t-t^\prime)$, where $n_T$ is the average number of thermal excitations at temperature $T$. Integration of (\ref{ptgamma}) yields
\begin{equation}\label{ptint}
\hat P(t)=\hat P(0)e^{-\gamma t/2}+{\cal P} (t)
+\int_0^t dt^\prime \hat f_\gamma(t^\prime)e^{\gamma (t^\prime-t)/2}\,,
\end{equation}
where ${\cal P}(t)=(\omega F/\gamma)[1-\exp(-\gamma t/2)]$  is the effective momentum displacement  produced by the force $F$.  This shows that, in the presence of friction, the momentum displacement due to the applied force is attenuated by $2[1-\exp(-\gamma t/2)]/\gamma t$ with respect to the noiseless displacement $\omega Ft/2$, obtained when $\gamma\rightarrow0$.

From (\ref{ptint}),  the evolution of the momentum variance is
\begin{equation}\label{varp}
\langle(\Delta\hat P)^2\rangle_t=\eta\langle(\Delta\hat P)^2\rangle_0+(2n_T+1)(1-\eta)/2\,, 
\end{equation}
where the index $t$ stands for the value of the variance at time $t$,  and $\eta\equiv\exp(-\gamma t)$.

If the initial state of the harmonic oscillator is Gaussian, so is the final state $\hat\rho_t$, since the oscillator is interacting with a thermal reservoir and the displacement is a Gaussian operation. Then the Fisher information for the force estimation corresponding to a momentum measurement on the final state of the oscillator is
\begin{equation}\label{fp}
{\cal F}_P(F)=\!\!\int \! dP{1\over\langle P|\hat\rho_t|P\rangle}\left({\partial\langle P|\hat\rho_t |P\rangle\over\partial F}\right)^2
={D^2(\eta)\over \langle(\Delta\hat P)^2\rangle_ t}\,,
\end{equation}
where $D(\eta)=(\omega/\gamma)(1-\sqrt{\eta})$ and $|P\rangle$ is an eigenstante of the momentum operator. Eqs. (\ref{varp}) and (\ref{fp}) imply that, for an initial minimum-uncertainty state in $X$ and $P$, 
\begin{equation}\label{fpx}
{\cal F}_P(F)=D^2(\eta){4\langle(\Delta\hat X)^2\rangle_0\over\eta+2(2n_T+1)(1-\eta)\langle(\Delta \hat X)^2\rangle_ 0}\,,
\end{equation}
where $\langle(\Delta\hat X)^2\rangle_ 0=1/[4\langle(\Delta\hat P)^2\rangle_0]$ is the variance of  $\hat X$ in the initial state. 

We show now that this expression coincides with the quantum Fisher information for the class of initial states considered above, and that no other class of states yields a larger quantum Fisher information for the estimation of the force. This is accomplished by applying the method proposed in Refs.~\cite{escher,escherbjp,nicim}. There it was shown that,  by transforming the non-unitary evolution of a noisy system into a unitary one in an extended space, it is possible to find an upper bound to the quantum Fisher information, which corresponds to the quantum Fisher information for the extended system (system plus environment). The idea is that measurements on both the system and the environment should not lead to less information about the parameter than measurements on the system alone.  It was shown that this bound can actually be attained, for some family of extensions. In this case, measurements on system plus environment yield redundant information with respect to measurements on the system alone. 

We discuss here in detail the case $T=0$,  which admits a simple physical picture. For $T\not=0$, the discussion follows similar lines, and is worked out in the Supplemental Material. 

\begin{figure}[t]
\centering
\includegraphics[width=0.43\textwidth]{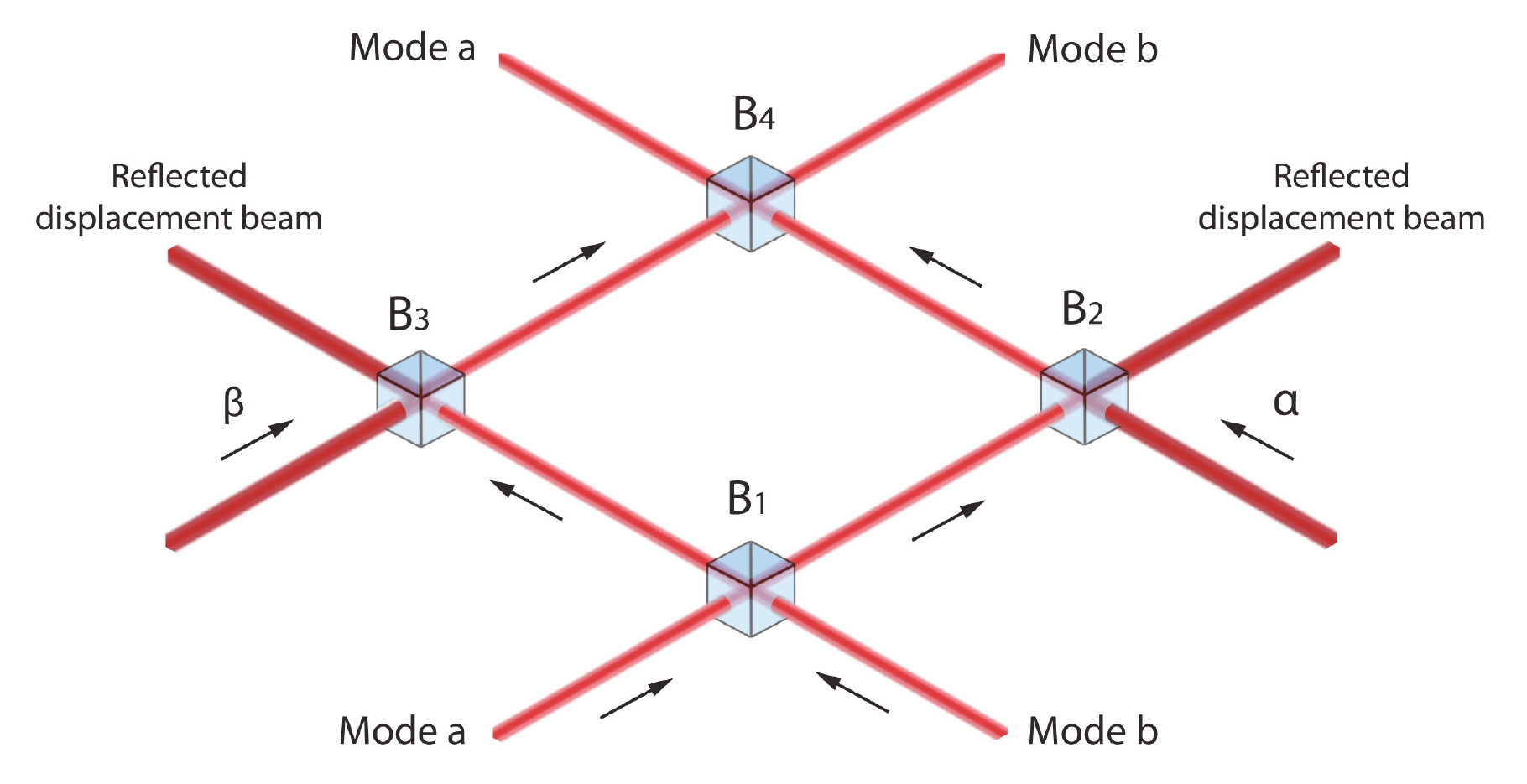}
\caption{Beam-splitter model for the coupling with the environment. The incoming beam, in mode {\bf a},  corresponds to the harmonic oscillator. Beam splitter $B_1$ induces photon losses in the incoming beam, as it deflects photons into mode {\bf b}, which corresponds to the environment.  Beam-splitter $B_2$, with transmissivity ${\cal T}\rightarrow0$, is used to displace the field in mode {\bf b}, through the injection of a high-intensity coherent state with amplitude $\alpha$, such that the product $\alpha\sqrt{{\cal T}}$ is finite. Beam splitter $B_3$ is used, analogously to $B_2$, to displace the field in the environment, upon injection of a coherent state with amplitude $\beta$. Beam splitter $B_4$ decouples modes {\bf a} and {\bf b}, allowing individual measurements on each mode.}
\label{bs}
\end{figure}

{\it Quantum Fisher information for the noisy oscillator.} Let us consider the harmonic oscillator $S$ as a mode of an electromagnetic field. The dissipation due to the interaction with the environment $R$ is modeled by the beam-splitter $B_1$, shown in Fig. \ref{bs}. This device deflects photons into mode {\bf b} (which plays the role of $R$).  If the transmissivity of this beam-splitter is $\eta=\exp(-\gamma t)$, then the evolution of the variances of the quadratures of the electromagnetic field is precisely that given by (\ref{varp}), when $n_T=0$, which  motivates this beam-splitter picture of the dissipation process. The displacement is induced by  a coherent state sent into a second beam splitter $B_2$, with  transmissivity ${\cal T}$ going to zero at the same time that the amplitude $\alpha$ of the coherent state goes to infinity, the product $\sqrt{{\cal T}}\alpha$ remaining finite. In the Supplemental Material, it is shown that these two operations yield an evolution for $S$ alone equivalent to the one derived from the master equation. The initial state of $S+R$ is $|\psi_0\rangle_S|0\rangle_R$,  the environment (mode {\bf b}) being initially in the vacuum state. 

After these two operations, the two-mode state becomes $ |\psi( F, t) \rangle_{SR} = \hat U_{SR}(F,t)|\psi_0 \rangle_S |0\rangle_R = e^{-i F  D(\eta) \hat{X}_S} \hat{B}_1 |\psi_0\rangle_S|0\rangle_R$, where $\hat{B}_1= e^{\theta_1(t)( \hat{a}\hat{b}^{\dag} - \hat{a}^{\dag}\hat{b})}$ is the beam-splitter operator  acting on modes {\bf a} and {\bf b}, with $\cos\theta_1(t) = \sqrt{\eta}$; $\hat{a}$ and $\hat{b}$ are the annihilation operators corresponding to modes {\bf a} and {\bf b}, respectively; $\hat{X}_S$ is a quadrature operator for mode {\bf a}, which corresponds to the position operator for the harmonic oscillator. The $\hat X_S$-dependent exponential in the above equation displaces the momentum as in \eqref{ptint}. 

In general, measurements on this extended system yield more information on the parameter than measurements on $S$ alone.  In order to reduce the non-redundant information about $F$ in $S+R$ we  displace the field in mode {\bf b}, along the same quadrature as the one displaced in mode {\bf a}.  This is implemented, as shown in Fig.~\ref{bs}, by sending a coherent state with amplitude $\beta$ on beam-splitter $B_3$, which has a vanishingly small transmittance, as was the case for $B_2$. This is a local operation on $R$, which does not affect $S$.  The evolution operator in $S+R$ takes then the form: 
\begin{equation}\label{usr}
\hat U_{SR}(G, t) = e^{i  FG(\eta) \hat{X}_R} e^{-i F D(\eta) \hat{X}_S} \hat B_1  \,,
\end{equation}
 where $\hat X_R$ is the position quadrature corresponding to mode {\bf b}.
 
Inserting \eqref{usr} into (\ref{h}) and (\ref{fq}), we get the respective quantum Fisher information: 
\begin{eqnarray}\label{fqg}
{\cal  F}^{SR}_Q( G)\!\! &=&\!\!\!  [-G(\eta)\sqrt{1-\eta} +  D(\eta)\sqrt{\eta}]^24 \langle (\Delta\hat X_S)^2 \rangle_0 \nonumber\\
 &+&\!\!\!  [D(\eta)\sqrt{1-\eta} + G(\eta)\sqrt{\eta}]^2 4 \langle (\Delta\hat X_R)^2 \rangle_0,
\end{eqnarray}
where the averages are calculated in the initial state of $S+R$. According to the above discussion, this should be an upper bound for the quantum Fisher information associated to $S$ alone for any value of $G$. The best upper bound is obtained by determining the function $G(\eta)$ that minimizes (\ref{fqg}). This is done in the Supplemental Material.

The corresponding minimum quantum Fisher information for $S+R$ coincides then precisely with  (\ref{fpx}) for a zero-temperature reservoir, with $\langle (\Delta\hat X)^2 \rangle_0 \equiv\langle (\Delta\hat X_S)^2 \rangle_0$.  The generalization of the above procedure for $T\not=0$, involving the addition of a squeezing transformation on mode {\bf a} and another environment mode {\bf c}, yields precisely Eq.~(\ref{fpx}) -- see Supplemental Material. This shows that the upper bound obtained by this minimization procedure is actually attained for initial minimum-uncertainty states in $X$ and $P$,  implying that it is the quantum Fisher information of $S$ alone, for these initial states.  It also implies that, for these states, the best measurement for the estimation of the force is a momentum measurement. 

Using \eqref{cr} and the upper bound obtained by minimizing \eqref{fqg}, one gets,  for any initial state of the harmonic oscillator with variance in $\hat X$ equal to  $\langle(\Delta\hat X)^2\rangle_0$, the following inequality for the uncertainty $\delta F$ in the estimation of the force:
\begin{equation}\label{dfnoise}
\delta F\ge {1\over D(\eta)\sqrt{2\nu}}\sqrt{(1-\eta)(2n_T+1)+{\eta/2\over \langle(\Delta\hat X)^2\rangle_0}}\,,
\end{equation}
this limit being attainable, for any integer value of $\nu$,  by Gaussian minimum-uncertainty states in $\hat X$ and $\hat P$. The standard limit corresponds to $\langle(\Delta\hat X)^2\rangle_0=1/2$, so that the right-hand side of Eq.~(\ref{dfnoise}) becomes $[D(\eta)\sqrt{2\nu}]^{-1}[2n_T(1-\eta)+1]^{1/2}$.

Since (\ref{dfnoise}) is a monotonous function of the variance $\langle (\Delta \hat X)^2 \rangle _0$,  one must maximize this quantity in order to minimize the lower bound on $\delta F$. For fixed average initial energy, this is achieved by a momentum-squeezed ground state of the harmonic oscillator, as shown in the Supplemental Material, for which $\langle (\Delta \hat X)^2 \rangle _0=E+\sqrt{E^2-1/4}$, where $E$ is the average energy in units of $\hbar\omega$. Since the dimensioneless force $F$ is related to the actual force $f$ by $F=f/\sqrt{\hbar m\omega^3}$, this implies that 
\begin{equation}\label{dfnoisefinal}
\delta f\ge{\sqrt{m\hbar\omega^3}\over  D(\eta)\sqrt{2\nu}}
\sqrt{(1-\eta)(2n_T+1)+{\eta/2\over E+\sqrt{E^2-1/4}}} .
\end{equation}
The equality is attained through momentum measurements on the oscillator, initialized in a momentum-squeezed ground state, and using the maximum-likelihood estimator \cite{Cramer1946,rao,fisher2}.

The lossless case is obtained for  $\gamma\rightarrow0$; then, only the second term inside the brackets is left. On the other hand, when $E+\sqrt{E^2-1/4}\gg \eta/[(1-\eta)(2n_T+1)]$ the lower bound becomes $\sqrt{m\hbar\omega^3}[D(\eta)\sqrt{2\nu}]^{-1}\sqrt{(1-\eta)(1+2n_T)}$. Therefore, for sufficiently large energy, one gets an expression similar to the one for a coherent state (standard limit), no matter how small are the losses, but lowered by a factor $\sqrt{(1-\eta)(2n_T+1)/[2n_T(1-\eta)+1]}$, which becomes $\sqrt{1-\eta}$ when $n_T\ll1$. 

\begin{figure}[t]
\centering
\includegraphics[width=0.32\textwidth]{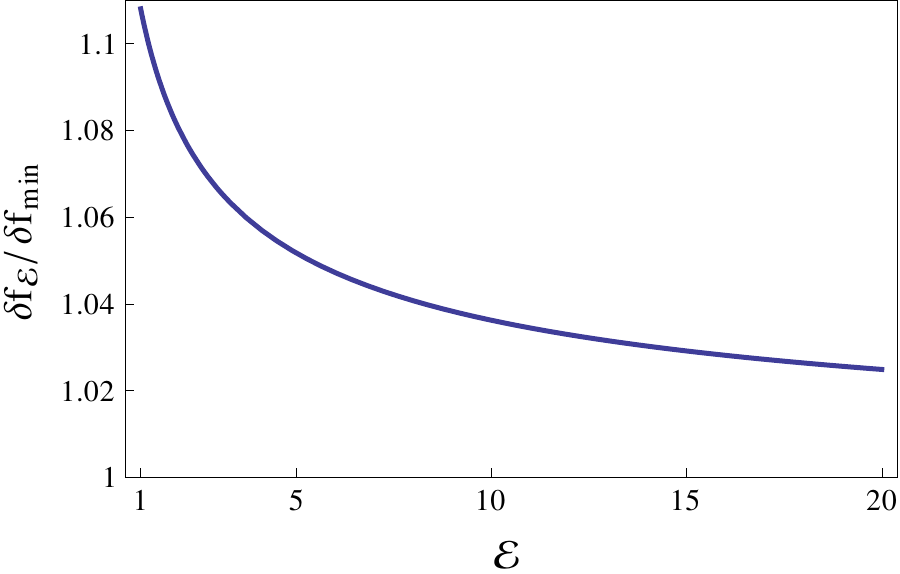}
\caption{Ratio between the energy- and temperature-dependent lower bound for the estimation of the force, obtained by optimizing the probing time,  and the strength of the thermal fluctuation, $\delta f_{\rm min}$, as a function of ${\cal E}=2\langle(\Delta\hat{X})^{2}\rangle_{0}(2n_{T}+1)$.}
\label{bound}
\end{figure}

{\it Sequential measurement procedure.} If the harmonic oscillator senses the force during a time $t$ such that $\gamma t\gg1$ and then is submitted to a single measurement, it follows from Eq.~(\ref{dfnoisefinal}) that the uncertainty in the estimation of the force will be limited by $\delta f_\infty\equiv\sqrt{m\hbar\omega\gamma^{2}{\rm cotgh}[\hbar\omega/(2k_{B}T)]/2}$, 
which is independent of the total probing time. This is due to the fact that the steady-state of the forced probe oscillator, which is interacting with a thermal reservoir, is independent of  its  initial state. It suggests that, if the unknown force acts during a time $t_{\rm tot}$, it is better to probe the force for an appropriate time $\tau$, measure the probe system, reset the system and repeat this procedure $\nu$ times, with $\nu=t_{\rm tot}/\tau$. With this approach, the total probing time is divided into $\nu$ parts  and the uncertainty is limited by
\begin{equation}\label{dftopt}
\delta f\ge\delta f_{\infty}\sqrt{\frac{1-(1-1/{\cal E})e^{-\gamma\tau}}{\nu(1-e^{-\gamma\tau/2})^2}} \ ,
\end{equation}
where ${\cal E}\equiv2\langle(\Delta\hat{X})^{2}\rangle_{0}(2n_{T}+1)$.

The optimal probing time $\tau_{\rm opt}$ leads to the minimum lower bound in Eq.~(\ref{dftopt}),  denoted by $\delta f_{\cal E}$. Then, for ${\cal E}\gg1$, the strategy of sequential and discrete measurements, performed in  the optimal dimensionless time $\gamma\tau_{\rm opt}\simeq[{\cal E}/24]^{-1/3}$ , leads to
\begin{equation}\label{bound_thermal}
\delta f_{\cal E}\ge\delta f_{\rm min}[1+(1/8) (3/{\cal E})^{2/3}+{\mathcal O}({\cal E}^{-1})]\,,
\end{equation} 
where $\delta f_{\rm min}\equiv\sqrt{(2m\hbar\omega\gamma/ t_{\rm tot}){\rm cotgh}\left(\hbar\omega/ 2k_{B}T\right)}$ is the probe ``potential sensitivity''  derived in Ref.~\cite{brag},  which measures the strength of the thermal fluctuation force acting on the oscillator and defines the maximum level of sensitivity in the estimation of  $f$,  valid for any measurement strategy. We have thus proven here that this bound is actually attainable asymptotically by a minimum-uncertainty squeezed state. 

The behavior of $\delta f_{\cal E}$ as a function of ${\cal E}$ is displayed in Fig.~\ref{bound}: for fixed temperature,  the gain obtained by squeezing the initial state  is  at most of the order of 10\%.

These results can also be applied to the diffusive situation considered in Ref. \cite{maiwald}, by letting $\gamma\rightarrow0$, $n_T\rightarrow\infty$, with $\gamma n_T={\cal D}$ (constant). One gets then from (\ref{dftopt}) that 
\begin{equation}
\delta f\ge \sqrt{{4m\hbar\omega {\cal D}/ t_{\rm tot}}}\sqrt{1+[4{\cal D}\langle(\Delta \hat X)^2\rangle_0 t_{\rm tot}]^{-1}}\,.
\end{equation}
In this diffusion limit, the optimal measurement time is the total time, which is a consequence of the fact that, while the displacement grows linearly with time, the noise in the momentum measurement  increases with the square root ot time. The first factor on the right-hand side of the above equation corresponds precisely to the expression derived through heuristic arguments in \cite{maiwald}. It is seen to overestimate the reachable precision, unless $4{\cal D}\langle(\Delta \hat X)^2\rangle_0 t_{\rm tot}\gg 1$. For the conditions assumed in \cite{maiwald}, this corresponds to $t_{\rm tot}\gg 1$ ms. 

In conclusion, we have been able to completely solve a precision problem involving the estimation of the amplitude of a resonant classical force through measurements on a probe consisting of a noisy quantum-mechanical oscillator. The force is estimated through a discrete sequence of measurements on the oscillator, at optimal time intervals. We have determined the ultimate precision limit, as a function of the average energy of the oscillator, and also the best  probe state and the best measurement procedure, thus yielding a rigorous and useful benchmark for experiments that aim to detect extremely small forces and displacements.

\begin{widetext}

\section{Supplemental Material}

\subsection{Evolution of the forced harmonic oscillator under a zero-temperature reservoir}\label{zero}

In this Section we derive some useful results concerning the evolution of the forced harmonic oscillator interacting with a zero-temperature reservoir. 

\subsubsection{Two-step decomposition of the evolution}

Here we show that the simultaneous actions of the external force and the noise can be decomposed into two successive operations. The first one  being a purely  dissipative evolution, corresponding to an interaction of the system with a zero-temperature thermal reservoir, and the second one corresponding to  a displacement in phase-space. 
 
 In the Markov limit, the master equation that describes the evolution of a harmonic oscillator in the presence of a resonant  force and a zero-temperature reservoir, in the interaction picture and under rotating-wave approximation,  is given by
\begin{equation}
\dfrac{d\hat{\rho}(t)}{dt}=-i\dfrac{\omega F}{2}\left[\hat{X},\hat\rho(t)\right]-\dfrac{\gamma}{2}\left[\hat{a}^{\dagger}\hat{a}\hat\rho(t)+\hat\rho(t)\hat{a}^{\dagger}\hat{a}-2\hat{a}\hat\rho(t)\hat{a}^{\dagger}\right] \ .\label{eq.lindblad}
\end{equation}
The unitary contribution on the right-hand side can be eliminated by choosing a convenient picture, where the density operator becomes
\begin{equation}
\hat\rho_{D}(t)=e^{i D(t)F \hat{X}}\hat{\rho}(t)e^{-i D(t)F\hat{X}} \ ,
\end{equation}
with $D(t)$ being a time-dependent function to be determined. In this picture, the  master equation becomes
\begin{equation}
\dfrac{d\hat\rho_{D}(t)}{dt}=i F\left(\dfrac{dD(t)}{dt}- \omega/2 +\dfrac{\gamma}{2}D(t)\right)\left[\hat{X},\hat\rho_{D}(t)\right]-\dfrac{\gamma}{2}\left[\hat{a}^{\dagger}\hat{a}\hat\rho_{D}(t)+\hat\rho_{D}(t)\hat{a}^{\dagger}\hat{a}-2\hat{a}\hat\rho_{D}(t)\hat{a}^{\dagger}\right] \ .
\end{equation}
The function $D(t)$ may be chosen as a solution of the equation
\begin{equation}
\dfrac{dD(t)}{dt}+\dfrac{\gamma}{2}D(t)-\omega/2=0 \ ,
\end{equation}
with the initial condition $D(0)=0$, which leads to the solution 
\begin{equation}
D(t) =\dfrac{\omega}{\gamma}(1-e^{-\gamma t/2}) \ ,
\end{equation}
noted $D(\eta) =\dfrac{\omega}{\gamma}(1-\sqrt{\eta})$ in the following, with $\eta=e^{-\gamma t}$.\\
Therefore, in this picture, and for the chosen  $D(\eta)$, the master equation that describes the evolution of $\hat\rho_{D}(t)$ is 
\begin{equation}\label{simple}
\dfrac{d\hat\rho_{D}(t)}{dt}=-\dfrac{\gamma}{2}\left[\hat{a}^{\dagger}\hat{a}\hat\rho_{D}(t)+\hat\rho_{D}(t)\hat{a}^{\dagger}\hat{a}-2\hat{a}\hat\rho_{D}(t)\hat{a}^{\dagger}\right] \ ,
\end{equation}
which only contains the effects of the interaction between the oscillator and the zero-temperature reservoir. 

This implies that the evolution of the system under the joint influence of the applied force and the interaction with the environment can be decomposed into a first step, involving only a purely dissipative evolution, followed by an effective  $\gamma$-dependent phase-space displacement. 

\subsubsection{Equivalence between the extended unitary evolution and the master equation treatment}\label{equiv}

We show now that there is a convenient purification of the non-unitary evolution stemming from  the above master equation, which yields a tight bound for the quantum Fisher information corresponding to the estimation of the force acting on the harmonic oscillator. 

The solution of \eqref{simple} may be found, for instance, in Ref.~\cite{fan08}. Going back to the interaction picture, one can see that the harmonic oscillator will evolve as 
\begin{equation}
\hat\rho(t)=e^{-i D(\eta)F \hat{X}}\left[\sum_{n=0}^{\infty} \frac{(1-\eta)^n}{n!}\eta^{\hat a^\dagger\hat a/2}\hat a^n\hat\rho_0(\hat a^\dagger)^n \eta^{\hat a^\dagger\hat a/2}\right]e^{i D(\eta)F \hat{X}},
\label{eq.tzero}\end{equation}
where $\hat\rho_0=|\psi_0\rangle_{S\,S}\langle\psi_0|$, and $|\psi_0\rangle_S$ is  the initial state of system $S$.

This non-unitary evolution can be seen as  resulting  from an unitary evolution on an enlarged Hilbert space comprising the system and an environment, when this environment is not monitored. Here,  we represent the environment by another harmonic oscillator
 and show that the unitary interaction between the system and this environment is enough to implement the non-unitary evolution given in Eq.~\eqref{eq.tzero}.
 
Let then
\begin{eqnarray}
|\Psi(t)\rangle&=&e^{-i D(\eta)F \hat{X}_S}\sum_{n=0}^{\infty}\sqrt{\frac{(1-\eta)^n}{n!}}\eta^{\hat a^\dagger\hat a/2}\hat a^n|\psi_0\rangle_S|n\rangle_R\nonumber\\
&=&e^{-i D(\eta)F \hat{X}_S}\sum_{n=0}^{\infty}\frac{(1-\eta)^{n/2}}{n!}\eta^{\hat a^\dagger\hat a/2}\hat a^n(\hat b^\dagger)^n|\psi_0\rangle_S|0\rangle_R
\end{eqnarray}
be a purification of $\hat\rho(t)$, that is, if one traces out the environment in $|\Psi(t)\rangle\langle\Psi(t)|$, one is left with $\hat\rho(t)$. In the above equation,  $ |n\rangle_R$ are Fock-states of the environment R, and $\hat a$ ($\hat b$) is an annihilation operator corresponding to system S (environment R). The subindex $S$ has been added to the position operator corresponding to system $S$. Now, it is straightforward to show that $|\Psi(t)\rangle$ can be rewritten as
\begin{equation}
|\Psi(t)\rangle=e^{-i D(\eta)F \hat{X}_S}\hat B_1|\psi_0\rangle_S|0\rangle_R,
\end{equation}
where $\hat B_1=e^{\arccos{(\sqrt{\eta})}(\hat{a}\hat{b}^{\dag} -\hat{a}^{\dag}\hat{b})}$ can be seen as the transformation performed by a beam-splitter with transmissivity $\eta$ on the input modes represented by $\hat a$ and $\hat b$.  Since $|\Psi(t)\rangle$ is a purification of $\hat\rho(t)$, the evolution described by the master equation~(\ref{eq.lindblad}) can equivalently be modeled by a beam-slpitter-like unitary interaction between the system S and the environment $R$ (represented here by a harmonic oscillator), followed by a displacement in phase-space of the system S, when the environment $R$ is not monitored (is traced out).

\subsubsection{Minimization of the quantum Fisher information  of  system plus environment }\label{sec.mqf}
The quantum Fisher information corresponding to system $S$ plus environment  $R$ is given by
\begin{equation}
{\cal  F}^{SR}_{Q}(G)=[-G(\eta)\sqrt{1-\eta} +  D(\eta)\sqrt{\eta}]^2 4\langle(\Delta\hat{X}_S)^2\rangle_0 + [D(\eta)\sqrt{1-\eta} + G(\eta)\sqrt{\eta}]^2 4\langle(\Delta\hat{X}_R)^2\rangle_0.
\end{equation}
Minimization of this quantity over $G$ leads to
\begin{displaymath}
 \left[(1- \eta)\langle (\Delta \hat{X}_S)^2\rangle_0 + \eta\langle(\Delta\hat{X}_R)^2\rangle_0\right] G(\eta) + \sqrt{\eta(1-\eta)}\left[\langle(\Delta\hat{X}_R)^2\rangle_0 - \langle (\Delta \hat{X}_S)^2\rangle_0 \right]D(\eta)=0. 
\end{displaymath}
The minimum is, therefore,  reached for 
\begin{equation}
G_{\rm opt}(\eta) = \frac{\sqrt{\eta(1-\eta)}\left[\langle (\Delta \hat{X}_S)^2\rangle_0 - \langle(\Delta\hat{X}_R)^2\rangle_0\right]}{(1-\eta)\langle (\Delta \hat{X}_S)^2\rangle_0 + \eta\langle(\Delta\hat{X}_R)^2\rangle_0}D(\eta).
\end{equation}
Inserting  this quantity into the expression of ${\cal F}^{SR}_Q (G)$ and using $\langle(\Delta\hat{X}_R)^2\rangle_0 = \frac{1}{2}$, since the environment $R$ is initially in the vacuum state, we obtain 
\begin{displaymath}
{\cal F}^{SR}_Q(G_{\rm opt}) =  D^2(\eta) \frac{4 \langle (\Delta \hat{X}_S)^2\rangle_0 }{\eta + 2(1-\eta)\langle (\Delta \hat{X}_S)^2\rangle_0 }.
\end{displaymath}

\subsection{Evolution of the forced harmonic oscillator under a thermal reservoir with $\mathbf{T\not=0}$}\label{evT}
 
In this Section, we generalize the previous results to the case of non-zero temperatures.

\subsubsection{Two-step decomposition of the evolution}

Here we show that the simultaneous actions of the external force and the thermal noise can also be decomposed into two successive operations.

When the probing harmonic oscillator interacts with a thermal reservoir at a temperature $T$, its evolution, in the interaction picture,  is dictated by the master equation 
\begin{eqnarray}\label{rhoT}
\dfrac{d\hat{\rho}(t)}{dt}&=&-i\dfrac{\omega F}{2}\left[\hat{X},\hat\rho(t)\right]-\dfrac{\gamma}{2}\left(n_{T}+1\right)\left[\hat{a}^{\dagger}\hat{a}\hat\rho(t)+\hat\rho(t)\hat{a}^{\dagger}\hat{a}-2\hat{a}\hat\rho(t)\hat{a}^{\dagger}\right] \nonumber\\
&-& \dfrac{\gamma n_{T}}{2}\left[\hat{a}\hat{a}^{\dagger}\hat\rho(t)+\hat\rho(t)\hat{a}\hat{a}^{\dagger}-2\hat{a}^{\dagger}\hat\rho(t)\hat{a}\right]  \ ,
\end{eqnarray}   
where $n_{T}$ is the mean number of thermal excitations in the reservoir, which obeys the Bose-Einstein distribution, given by
\begin{equation}
n_{T}=\dfrac{1}{e^{\hbar\omega/k_{B}T}-1} \ ,
\end{equation}
whith  $k_{B}$  being  the Boltzmann constant.

This master equation is simplified by using the same procedure as in the zero-temperature case, that is, one defines
\begin{equation}
\hat\rho_{D}(t)=e^{iD(\eta)F \hat{X}}\hat{\rho}(t)e^{-i D(\eta)F\hat{X}} \ ,
\end{equation}
with $D(\eta)$ defined as before. One finds then that $\hat \rho_D$ satisfies the above master equation without the force term.

\subsubsection{Equivalence between the extended unitary evolution and the Master equation treatment}\label{equivT}

The solution of \eqref{rhoT} without the force term may be found in Ref.~\cite{fan08}:
\begin{equation}
\hat\rho_{D}(t)=r_{3}e^{\ln{[r_{2}]}\hat{a}^{\dagger}\hat{a}}\sum_{l,j=0}^{\infty}\left[\dfrac{(n_{T}+1)^{l}(n_{T})^{j}r_{1}^{l+j}}{l!j!r_{2}^{2j}}(\hat{a}^{\dagger})^{j}\hat{a}^l\hat\rho_{0}(\hat{a}^\dagger)^l\hat{a}^{j}\right]e^{\ln{[r_{2}]}\hat{a}^{\dagger}\hat{a}} \ ,
\end{equation}
where the functions $r_{1}$, $r_{2}$ and $r_{3}$ are defined by
\begin{eqnarray*}
r_{1}&=&\dfrac{1-\eta}{\left[n_{T}(1-\eta)+1\right]} \ ,\\
r_{2}&=&\dfrac{\sqrt{\eta}}{n_{T}(1-\eta)+1} \ , \\
r_{3}&=&\dfrac{1}{n_{T}(1-\eta)+1} \ ,
\end{eqnarray*} 
and $\eta=e^{-\gamma t}$. Therefore, in the interaction picture, the evolution of $\hat\rho(t)$ is
\begin{equation}
\hat\rho(t)=e^{-i D(\eta)F \hat{X}}\left\{r_{3}e^{\ln{[r_{2}]}\hat{a}^{\dagger}\hat{a}}\sum_{l,j=0}^{\infty}\left[\dfrac{(n_{T}+1)^{l}(n_{T})^{j}r_{1}^{l+j}}{l!j!r_{2}^{2j}}(\hat{a}^{\dagger})^{j}\hat{a}^l\hat\rho_{0}(\hat{a}^\dagger)^l\hat{a}^{j}\right]e^{\ln{[r_{2}]}\hat{a}^{\dagger}\hat{a}}\right\}e^{i D(\eta)F \hat{X}} \ .
\end{equation}

A purification of $\hat\rho(t)$ can be built from this solution. Notice that, to find this purification, one considers an environment with two harmonic oscillators. Such a  purification may  be given by
\begin{eqnarray*}
\vert\Psi(t)\rangle&=&e^{-i D(\eta)F \hat{X}_S}e^{\ln{[r_{2}]}\hat{a}^{\dagger}\hat{a}}\sum_{l,j=0}^{\infty}\sqrt{\dfrac{(n_{T}+1)^{l}(n_{T})^{j}r_{1}^{l+j}r_{3}}{l!j!r_{2}^{2j}}}(\hat{a}^{\dagger})^{j}\hat{a}^l\vert\psi_{0}\rangle_{S}\vert l\rangle_{R_{1}}\vert j\rangle_{R_{2}} \\
&=&e^{-i D(\eta)F \hat{X}_S}e^{\ln{[r_{2}]}\hat{a}^{\dagger}\hat{a}}\sum_{l,j=0}^{\infty}\sqrt{\dfrac{(n_{T}+1)^{l}(n_{T})^{j}r_{1}^{l+j}r_{3}}{r_{2}^{2j}}}\dfrac{(\hat{a}^{\dagger}\hat{b}^{\dagger})^{j}}{j!}\dfrac{(\hat{a}\hat{c}^{\dagger})^l}{l!}\vert\psi_{0}\rangle_{S}\vert0\rangle_{R_{1}}\vert0\rangle_{R_{2}} \ ,
\end{eqnarray*}
where the states $\vert l\rangle_{R_{1}}$ and $\vert j\rangle_{R_{2}}$ are Fock states of the environments $R_{1}$ and $R_{2}$ respectively, and $\hat{b}$ ($\hat{c}$) is the annihilation operator for the environment $R_{1}$ ($R_{2}$). The subindex $S$ has been added to denote an operator acting on $S$. This purification may be seen in terms of three unitary evolutions: the first one corresponds to a beam-splitter-like interaction between the system $S$ and the environment $R_{1}$, the second one corresponds to a two-mode squeezing-like interaction between the system $S$ and the environment $R_{2}$, and the third one corresponds to phase-space displacement in $S$ space. Defining the  first two  unitary operators by
\begin{eqnarray*}
\hat{B}_1&=&e^{\theta_{1}(t)\left(\hat{a}\hat{b}^{\dagger}-\hat{a}^\dagger\hat{b}\right)} \ , \\
\hat{S}&=&e^{\theta_{2}(t)\left(\hat{a}^{\dagger}\hat{c}^{\dagger}-\hat{a}\hat{c}\right)} \ , 
\end{eqnarray*}
where $\theta_{1}(t)$ and $\theta_{2}(t)$ are given by
\begin{eqnarray*}
\theta_{1}(t)&=& {\rm arccos}\left[\sqrt{\dfrac{\eta}{n_{T}(1-\eta)+1}}\right] \ , \\
\theta_{2}(t)&=& {\rm arccosh}\left[\sqrt{n_{T}(1-\eta)+1}\right] ,
\end{eqnarray*}
then, the above purification can be rewritten as
\begin{equation}
\vert\Psi(t)\rangle=e^{-i D(\eta)F \hat{X}_S}\hat{S}\hat{B}_1\vert\psi_{0}\rangle_{S}\vert0\rangle_{R_{1}}\vert0\rangle_{R_{2}} \ .
\end{equation}

\subsubsection{Minimization of the quantum Fisher information of system plus environment}

One should note that there are many purifications of the state $\hat\rho(t)$ of $S$, corresponding to different states of $S+R_1+R_2$.  Indeed, for any Hermitean operator $\hat{H}_{1,2}$ acting only on the environments $R_{1}$ and $R_{2}$, the pure state $\vert\Phi\rangle$, given by
\begin{equation}
\vert\Phi\rangle=e^{-i D(\eta)F\hat{H}_{1,2}}\vert\Psi(t)\rangle \ ,
\end{equation}
is another purification of $\hat\rho(x)$. Notice also that $\hat{H}_{1,2}$ should be chosen properly in order to minimize the quantum Fisher information in $S+R_{1}+R_{2}$.

An upper bound to the quantum Fisher information in $S$ may be calculated with $\vert\Phi\rangle$. This bound is given by
\begin{equation}\label{fqpsi}
{\mathcal F}^{SR_1R_2}_{Q}=\left[2 D(\eta)\right]^{2}_{ \ R_{1}}\!\!\langle0\vert{\!\!\!\phantom]}_{\ R_{2}}\!\langle0\vert{\!\phantom]}_{S} \langle\psi_{0}\vert\left\{\Delta\left[ \hat{B}_1^{\dagger}\hat{S}^\dagger\left(\hat{X}_{S}+\hat{H}_{1,2}\right)\hat{S}\hat{B}_1\right] \right\}^{2}\vert\psi_{0}\rangle_S\vert0\rangle_{R_{2}}\vert0\rangle_{R_{1}} \ .
\end{equation}
A possible choice of the operator $\hat{H}_{1,2}$ aimed to erase part of the non-redundant information about the value of the force $F$ in $\vert\Phi\rangle$, as compared with the same information in $\hat\rho(t)$, is $\hat{H}_{1,2}=\lambda_{1}\hat{X}_{R_{1}}+\lambda_{2}\hat{X}_{R_{2}}$, where $\hat{X}_{R_{1}}$ ($\hat{X}_{R_{2}}$) is the position operator of the oscillator in $R_{1}$ ($R_{2}$) space. This choice is based on physical insights on the enlarged unitary process. Since $\vert\Phi\rangle$ is an entangled state in $S+R_{1}+R_{2}$, the effect of a phase-space displacement along the momentum-axis in $S$ may affect the state in $S+R_{1}+R_{2}$, by changing the correlations between the system and the environments. Indeed, for $\hat{H}_{1,2}=0$, after disentangling $S+R_{1}+R_{2}$ with the operation $\hat{B}_1^{\dagger}\hat{S}^\dagger$, which does not change the quantum Fisher information, the effective unitary evolution in $S+R_{1}+R_{2}$ is
\begin{equation}
\hat{U}_{S,R_{1},R_{2}}=\exp\left\{-i D(\eta)F \left[\cosh\left[\theta_{2}(t)\right]\cos\left[\theta_{1}(t)\right]\hat{X}_S-\cosh\left[\theta_{2}(t)\right]\sin\left[\theta_{1}(t)\right]\hat{X}_{R_{1}}+\sinh\left[\theta_{2}(t)\right]\hat{X}_{R_{2}}\right]\right\} \ .
\end{equation}
It is clear that, with $\hat{H}_{1,2}=\lambda_{1}\hat{X}_{R_{1}}+\lambda_{2}\hat{X}_{R_{2}}$ and convenient values of $\lambda_{1}$ and $\lambda_{2}$, it is possible to erase at least part of the non-redundant information in $\vert\Psi(t)\rangle$.

After a straightforward calculation, Eq.~(\ref{fqpsi}), with the above choice of $\hat{H}_{1,2}$, can be rewritten as
\begin{eqnarray}
\dfrac{{\mathcal F}^{SR_1R_2}_{Q}(\lambda_{1},\lambda_{2})}{\left[2 D(\eta)\right]^{2}}&=&\langle(\Delta\hat{X}_S)^{2}\rangle_{0}\left[(\cosh[\theta_{2}(t)]+\lambda_{2}\sinh[\theta_{2}(t)])\cos[\theta_{1}(t)]+\lambda_{1}\sin[\theta_{1}(t)]\!\!\!\!\phantom{\frac{1}{1}}\right]^{2}\nonumber\\
&+&\langle(\Delta\hat{X}_{R_{1}})^{2}\rangle_{0}\left[-(\cosh[\theta_{2}(t)]+\lambda_{2}\sinh[\theta_{2}(t)])\sin[\theta_{1}(t)]+\lambda_{1}\cos[\theta_{1}(t)]\!\!\!\!\phantom{\frac{1}{1}}\right]^{2}\nonumber\\
&+&\langle(\Delta\hat{X}_{R_{2}})^{2}\rangle_{0}\left[\sinh[\theta_{2}(t)]+\lambda_{2}\cosh[\theta_{2}(t)]\!\!\!\!\phantom{\frac{1}{1}}\right]^{2} \ .
\end{eqnarray}
Then, the optimal values of $\lambda_{1}$ and $\lambda_{2}$, which minimize the above equation, are
\begin{eqnarray*}
\lambda_{1}^{\rm (opt)}&=&\left[\dfrac{(\langle(\Delta\hat{X}_{R_{1}})^{2}\rangle_{0}-\langle(\Delta\hat{X}_S)^{2}\rangle_{0})\cos[\theta_{1}(t)]\sin[\theta_{1}(t)]}{\langle(\Delta\hat{X}_{R_{1}})^{2}\rangle_{0}\cos^{2}[\theta_{1}(t)]+\langle(\Delta\hat{X}_S)^{2}\rangle_{0}\sin^{2}[\theta_{1}(t)]}\right](\cosh[\theta_{2}(t)]+\lambda_{2}^{\rm(opt)}\sinh[\theta_{2}(t)]) \\
\lambda_{2}^{\rm (opt)}&=&-\cosh[\theta_{2}(t)]\sinh[\theta_{2}(t)]\nonumber\\
&\times&\left[\dfrac{\langle(\Delta\hat{X}_{R_{2}})^{2}\rangle_{0}+\dfrac{\langle(\Delta\hat{X}_S)^{2}\rangle_{0}\langle(\Delta\hat{X}_{R_{1}})^{2}\rangle_{0}}{\langle(\Delta\hat{X}_{R_{1}})^{2}\rangle_{0}\cos^{2}[\theta_{1}(t)]+\langle(\Delta\hat{X}_S)^{2}\rangle_{0}\sin^{2}[\theta_{1}(t)]}}{\cosh^2[\theta_{2}(t)]\langle(\Delta\hat{X}_{R_{2}})^{2}\rangle_{0}+\sinh^{2}[\theta_{2}(t)]\dfrac{\langle(\Delta\hat{X}_S)^{2}\rangle_{0}\langle(\Delta\hat{X}_{R_{1}})^{2}\rangle_{0}}{\langle(\Delta\hat{X}_{R_{1}})^{2}\rangle_{0}\cos^{2}[\theta_{1}(t)]+\langle(\Delta\hat{X}_S)^{2}\rangle_{0}\sin^{2}[\theta_{1}(t)]}}\right] \ .
\end{eqnarray*}
Therefore, the minimum value of ${\mathcal F}^{SR_1R_2}_{Q}(\lambda_{1},\lambda_{2})$ is given by
\begin{equation}
\dfrac{{\mathcal F}^{SR_1R_2}_{Q}(\lambda_{1}^{\rm (opt)},\lambda_{2}^{\rm (opt)})}{\left[2D(\eta)\right]^{2}}=\left[\dfrac{\sinh^{2}[\theta_{2}(t)]}{\langle(\Delta\hat{X}_{R_{1}})^{2}\rangle_{0}}+\dfrac{\cosh^{2}[\theta_{2}(t)]\sin^{2}[\theta_{1}(t)]}{\langle(\Delta\hat{X}_{R_{2}})^{2}\rangle_{0}}+\dfrac{\cosh^{2}[\theta_{2}(t)]\cos^{2}[\theta_{1}(t)]}{\langle(\Delta\hat{X}_S)^{2}\rangle_{0}}\right]^{-1} \ .
\end{equation}
Substituting  $\theta_{1}(t)$ and $\theta_{2}(t)$ in terms of $\eta$ and $n_{T}$, and setting  $\langle(\Delta\hat{X}_{R_{1}})^{2}\rangle_{0}=\langle(\Delta\hat{X}_{R_{2}})^{2}\rangle_{0}=1/2$ in the above equation, since both environments start in a vacuum state of the harmonic oscillator,  we get the following upper bound
\begin{equation}
{\mathcal F}^{SR_1R_2}_{Q}(\lambda_{1}^{\rm (opt)},\lambda_{2}^{\rm (opt)})=[2 D(\eta)]^{2}\left[2(1-\eta)(2n_{T}+1)+\dfrac{\eta}{\langle(\Delta\hat{X}_S)^{2}\rangle_{0}}\right]^{-1} \ .
\end{equation}

\subsection{Maximization of $\langle  (\Delta \hat X)^2 \rangle_0 $}

We want to maximize the variance $\langle  (\Delta\hat  X)^2 \rangle_0$ under  the constraint of fixed resource $\langle\hat X^2\rangle_0 + \langle\hat P^2\rangle_0 = 2E$ and taking into account the physical restriction imposed by 
the Heisenberg uncertainty relation $\langle  (\Delta \hat X)^2 \rangle_0\langle ( \Delta \hat P)^2 \rangle_0 = a \geq 1/4$. This leads to a system of two equations with two unknown
parameters:
\begin{eqnarray}
 \langle ( \Delta \hat X)^2 \rangle_0\langle ( \Delta \hat P)^2 \rangle_0 &=& a\\
 \langle  (\Delta \hat X)^2 \rangle_0 + \langle ( \Delta \hat P)^2 \rangle_0 &=& 2 E^*
\end{eqnarray}
with $2E^* = 2E -\langle  \hat X\rangle_0^2 - \langle   \hat P \rangle_0^2$.
 
The solutions are $\langle  (\Delta \hat X)^2 \rangle_0^{\pm} = E^* \pm \sqrt{E^{*2} - a}$.
 It follows that the maximum  value of the variance in $\hat X$ under the  two constraints imposed above is 
$\langle ( \Delta \hat X)^2 \rangle_0 \equiv E^* + \sqrt{E^{*2} - 1/4}$, reached when $a=1/4$, that is, when the state is  a minimum uncertainty state in $X$ and $P$. 

Minimum-uncertainty states in $X$ and $P$ are pure Gaussian states, which can be obtained from the ground state through squeezing characterized by a real parameter $r$ and a complex displacement $z$. Because  the variances $\langle  (\Delta \hat X)^2 \rangle_0$ and $\langle  (\Delta \hat P)^2 \rangle_0$ do not depend on the value of  $z$, the best strategy to maximize the variance $\langle  (\Delta \hat X)^2 \rangle_0$, for a fixed value $E$ of the average energy, is to invest all the energy in the squeezing of the state. This allows one to conclude that  the state that maximizes the variance of $\hat X$, under the constraints imposed above, is a squeezed ground state
 $|\psi_0 \rangle = S(r) |0\rangle$, with $ S(r)  = \exp{\frac{r}{2}(\hat{a}^2 - \hat{a}^{\dag 2})} $ and  $r = 1/2 \ln{[2 ( E + \sqrt{E^{2} - 1/4})]}$.
\end{widetext}

The authors acknowledge the support of the Brazilian agencies CNPq, CAPES, FAPERJ, and the National Institute of Science and Technology for Quantum Information.

\end{document}